\documentstyle[prd,aps,eqsecnum,preprint,tighten,floats,epsf]{revtex}
\begin{document}
\draft

\preprint{\vbox{\hfill SLAC-PUB-8078 \\
          \vbox{\hfill UMN-D-99-1}  \\
          \vbox{\hfill SMUHEP/99-04 }
          \vbox{\vskip0.3in}
          }}

\title{Application of Pauli--Villars Regularization 
and Discretized Light-Cone Quantization 
to a ($3+1$)-Dimensional Model}

\author{Stanley J. Brodsky%
\footnote{\baselineskip=14pt
Work supported in part by the Department of Energy,
contract DE-AC03-76SF00515.}}
\address{Stanford Linear Accelerator Center,
Stanford University, Stanford, California 94309}

\author{John R. Hiller%
\footnote{\baselineskip=14pt
Work supported in part by the Department of Energy,
contract DE-FG02-98ER41087.}}
\address{Department of Physics,
University of Minnesota, Duluth, Minnesota 55812}

\author{Gary McCartor%
\footnote{\baselineskip=14pt
Work supported in part by the Department of Energy,
contract DE-FG03-95ER40908.}}
\address{Department of Physics,
Southern Methodist University, Dallas, Texas 75275}

\date{\today}

\maketitle

\begin{abstract}
We apply Pauli--Villars regularization and
discrete light-cone quantization to the nonperturbative
solution of a ($3+1$)-dimensional model field theory.
The matrix eigenvalue problem is solved for the
lowest-mass state with use of the complex symmetric
Lanczos algorithm.  This permits the calculation
of each Fock-sector wave function, and from these we
obtain values for various quantities, such as
average multiplicities and average momenta of
constituents, structure functions, and a
form factor slope.
\end{abstract}
\pacs{12.38.Lg,11.15.Tk,11.10.Gh,02.60.Nm
\begin{center}(Submitted to Physical Review D.)\end{center}}

\narrowtext

\section{Introduction}

One of the most challenging problems in particle physics is the
computation of the spectrum and physical properties of bound states in
quantum field theory. The main tool presently used for such
nonperturbative computations in quantum chromodynamics is lattice gauge
theory\cite{WilsonLGT}, which has been highly  successful for determining
hadron spectra. However, the computation of dynamical properties, such as
CP violation in weak transition matrix elements\cite{Neubert} or
the shape of the distributions measured in deep inelastic scattering is
difficult using standard lattice methods.

Light-cone Hamiltonian diagonalization methods\cite{DLCQreview}
appear to provide a
number of attractive advantages for solving nonperturbative problems in
quantum field theory, including a Minkowski space description, boost
invariance, no fermion-doubling, and a consistent Fock state expansion
well matched to physical problems in QCD; however, thus far,  full
dynamical solutions based on light-cone Hamiltonian diagonalization have
been primarily limited to one-space/one-time models. One promising
approach is the transverse lattice which combines light-cone methods in
the longitudinal light-cone direction with a spacetime lattice for the
transverse dimensions.\cite{Dalley}

In recent work\cite{PV1} we have shown that a model field
theory in $3+1$ dimensions can be solved using
discrete light-cone quantization
(DLCQ)\cite{PauliBrodsky,DLCQreview}, a light-cone Hamiltonian
diagonalization method, together with Pauli--Villars
regulation of the ultraviolet\cite{PauliVillars}. The particular model
theory which we constructed has an exact analytic solution by which the
DLCQ results could be checked, for both accuracy and rapidity of
convergence. The model was regulated in the ultraviolet by a single
Pauli--Villars boson, which was included in the DLCQ Fock basis in the
same way as the ``physical'' particles of the theory.  The two bare
parameters of the model were then determined by fits of observables to
chosen values.

Here we shall extend this combination of DLCQ and Pauli--Villars
regularization to a more realistic model which mimics many features
of a full quantum field theory. Unlike the analytic model which contained
a static source, the  light-cone energies of the particles in the new
model have the correct longitudinal and transverse momentum dependence.
Although an analytic solution of the new model is no longer available, the
numerical convergence of the discretized light-cone solutions is found to
be quite rapid, and the structure of the solution for the lowest-mass
eigenstate is readily obtained. In particular, we can calculate the
light-cone wavefunction of each Fock-sector component, and from these we
can compute the values for various physical quantities, such as average
multiplicities and average momenta of constituents, bosonic and fermionic
structure functions, and a form factor slope.

A distinct advantage of our approach is that almost
all counterterms are generated automatically by the
Pauli--Villars particles and their imaginary couplings.
This can be explicitly checked for consistency in perturbation theory. For
nonperturbative calculations we conjecture that the same number of
Pauli--Villars fields will be sufficient to regulate the theory.  This
does appear to be the case here and in the work reported
previously\cite{PV1}.  An alternative procedure has been proposed and
explored by Wilson, Perry and collaborators\cite{Wilson}; they use a
similarity transformation to generate effective Hamiltonians
perturbatively which can then be diagonalized in the valence Fock sector.

In our approach one can obtain the full set
of Fock-sector wave functions for the lowest-mass eigenstate.   This
contrasts with other DLCQ calculations in $3+1$
dimensions\cite{Hollenberg,TangPauli,ae} where
the number of particles was severely limited from the
outset and effects of higher Fock sectors can only
be estimated.  The DLCQ calculation
by Wivoda and Hiller\cite{Wivoda}, though untruncated,
did not construct counterterms in a way that can be
systematically extended to other theories.   In our case,
a Tamm--Dancoff truncation\cite{TammDancoff} in
particle number can be applied, and the impact of the
truncation can be studied and understood.

Our notation is such that we define light-cone
coordinates\cite{Dirac} by
\begin{equation}
x^\pm = x^0+x^3\,,\;\;
\bbox{x}_\perp=(x^1,x^2)\,.
\end{equation}
The time coordinate is taken to be $x^+$.  The dot
product of two four-vectors is
\begin{equation}
p\cdot x=\frac{1}{2}(p^+x^- + p^+x_-)
                -\bbox{p}_\perp\cdot\bbox{x}_\perp\,.
\end{equation}
Thus the momentum component conjugate to $x^-$ is $p^+$,
and the light-cone energy is $p^-$.  We use underscores
to identify light-cone three-vectors, such as
\begin{equation}
\underline{p}=(p^+,\bbox{p}_\perp)\,.
\end{equation}
For additional details, see Appendix A of Ref.~\cite{PV1}
or a review paper \cite{DLCQreview}.

The model which we study is defined in Sec.~\ref{sec:Model}.
There we also list and define various quantities which we will compute
from the eigensolution, including structure functions and distribution
amplitudes, average multiplicities, and average momenta.  The numerical
methods, including the DLCQ procedure, and the results are discussed in
Sec.~\ref{sec:NumMethods}.  Section~\ref{sec:Conclusions} contains some
concluding remarks and plans for future work.

\section{A Model with a Dynamical Source}  \label{sec:Model}


We shall consider a field-theoretic model where one particle, which
we take to be a fermion of mass $M$, acts as a dynamical source and sink
for bosons of mass $\mu$.  The model is only slightly more complicated
than the analytically soluble model considered in Ref.~\cite{PV1}, the key
difference being that here the fermion has a proper, momentum-dependent
light-cone energy.  Another difference is that the vertices do not include
the momentum ratios which were introduced in\cite{PV1} to control
end-point behavior; the restoration of fermion dynamics makes such factors
unnecessary.  The theory is still regulated by a single Pauli--Villars
boson with imaginary couplings\footnote{One could use an Hermitian form
and negative metric to implement Pauli--Villars regularization, but
the complex symmetric form is what is known to work well with the 
numerical method we have chosen.}
and a mass $\mu_1$.  The light-cone Hamiltonian (or
mass-squared operator) $H_{\rm LC}=P^+P^--P_\perp^2$ is,
in the $\bbox{P}_\perp=0$ frame,
\begin{eqnarray} 
\label{eq:ModelH} \lefteqn{H_{\rm LC}=
       \int\frac{dp^+d^2p_\perp}{16\pi^3p^+}
           \left(\frac{M^2+p_\perp^2}{p^+/P^+}+M'_0\frac{p^+}{P^+}\right)
                 \sum_\sigma b_{\underline{p}\sigma}^\dagger
                        b_{\underline{p}\sigma}} \hspace{0.5in} \\
  & +&\int\frac{dq^+d^2q_\perp}{16\pi^3q^+}
       \left[\frac{\mu^2+q_\perp^2}{q^+/P^+}
                       a_{\underline{q}}^\dagger a_{\underline{q}}
           + \frac{\mu_1^2+q_\perp^2}{q^+/P^+}
                        a_{1\underline{q}}^\dagger a_{1\underline{q}}
               \right]  \nonumber \\
    & +&g\int\frac{dp_1^+d^2p_{\perp1}}{\sqrt{16\pi^3p_1^+}}
            \int\frac{dp_2^+d^2p_{\perp2}}{\sqrt{16\pi^3p_2^+}}
              \int\frac{dq^+d^2q_\perp}{16\pi^3q^+}
                \sum_\sigma b_{\underline{p}_1\sigma}^\dagger
                             b_{\underline{p}_2\sigma}
   \nonumber \\
     & &\times \left[
      a_{\underline{q}}^\dagger
              \delta(\underline{p}_1-\underline{p}_2+\underline{q})
        +a_{\underline{q}}
              \delta(\underline{p}_1-\underline{p}_2-\underline{q})
              \right.
    \nonumber \\
     & & \left.
       +ia_{1\underline{q}}^\dagger
             \delta(\underline{p}_1-\underline{p}_2+\underline{q})
      +ia_{1\underline{q}}
            \delta(\underline{p}_1-\underline{p}_2-\underline{q})
            \right]\,,
    \nonumber
\end{eqnarray}
where $b_{\underline{p}\sigma}^\dagger$, $a_{\underline{q}}^\dagger$, and
$a_{1\underline{q}}^\dagger$ are creation operators for the fermion
source, the physical boson, and the Pauli--Villars boson, respectively. 
The operators obey the usual commutation relations 
\begin{eqnarray}
\label{eq:CommRelations}
\left\{b_{\underline{p}\sigma},b_{\underline{p}'\sigma'}^\dagger\right\}
     &=&16\pi^3p^+\delta(\underline{p}-\underline{p}')
                                     \delta_{\sigma\sigma'}\,,
  \nonumber \\
\left[a_{\underline{q}},a_{\underline{q}'}^\dagger\right]
          &=&16\pi^3q^+\delta(\underline{q}-\underline{q}')\,,
  \nonumber \\
\left[a_{1\underline{q}},a_{1\underline{q}'}^\dagger\right]
          &=&16\pi^3q^+\delta(\underline{q}-\underline{q}')\,.
\end{eqnarray}
The $M'_0p^+/P^+$ counterterm is inserted to cancel a
logarithmic dependence on the Pauli--Villars mass which
arises from the one-loop self-energy integral
\begin{equation}
 \frac{g^2}{16\pi^3}\left\{\int_0^{p^+}\frac{dq^+}{q^+}
    \frac{d^2q_\perp}
     {\frac{M^2+p_\perp^2}{p^+/P^+}
          -\frac{M^2+({\bf p}_\perp+{\bf q}_\perp)^2}{(p^+-q^+)/P^+}
                      -\frac{\mu^2+{\bf q}_\perp^2}{q^+/P^+}}
  -\mbox{P-V term}\right\}
     \sim -\frac{g^2}{16\pi^2}\ln(\mu_1/\mu)\,.
\end{equation}
This model Hamiltonian is distantly related to the Yukawa
Hamiltonian\cite{McCartorRobertson}, to which one might
also eventually apply the techniques used here.

The bare parameters $g$ and $M'_0$ are to be fixed by fitting
physical properties of the lowest massive eigenstate.  This
is a dressed fermion state which we write as
\begin{eqnarray}
\Phi_\sigma&=&\sqrt{16\pi^3P^+}\sum_{n,n_1}
                    \int\frac{dp^+d^2p_\perp}{\sqrt{16\pi^3p^+}}
   \prod_{i=1}^n\int\frac{dq_i^+d^2q_{\perp i}}{\sqrt{16\pi^3q_i^+}}
   \prod_{j=1}^{n_1}\int\frac{dr_j^+d^2r_{\perp j}}{\sqrt{16\pi^3r_j^+}}
   \\ 
    &  & \times \delta(\underline{P}-\underline{p}
                     -\sum_i^n\underline{q}_i-\sum_j^{n_1}\underline{r}_j)
       \phi^{(n,n_1)}(\underline{q}_i,\underline{r}_j;\underline{p})
         \frac{1}{\sqrt{n!n_1!}}b_{\underline{p}\sigma}^\dagger
          \prod_i^n a_{\underline{q}_i}^\dagger
             \prod_j^{n_1} a_{1\underline{r}_j}^\dagger |0\rangle \,,
   \nonumber
\end{eqnarray}
and normalize according to
\begin{equation}
\Phi_\sigma^{\prime\dagger}\cdot\Phi_\sigma
=16\pi^3P^+\delta(\underline{P}'-\underline{P})\,.
\end{equation}
The individual amplitudes must then satisfy
\begin{equation}  \label{eq:NormCondition}
\sum_{n,n_1}\prod_i^n\int\,dq_i^+d^2q_{\perp i}
                     \prod_j^{n_1}\int\,dr_j^+d^2r_{\perp j}
    \left|\phi^{(n,n_1)}(\underline{q}_i,\underline{r}_j;
           \underline{P}-\sum_i\underline{q}_i
                              -\sum_j\underline{r}_j)\right|^2=1\,.
\end{equation}


The eigenvalue problem is
\begin{equation}
H_{\rm LC}\Phi_\sigma=M^2\Phi_\sigma\,.
\end{equation}
This is equivalent to the following coupled set of integral equations for
the amplitudes: 
\begin{eqnarray} \label{eq:CoupledEqns}
\lefteqn{\left[M^2-\frac{M^2+p_\perp^2}{x}-M'_0x
  -\sum_i\frac{\mu^2+q_{\perp i}^2}{y_i}
                  -\sum_j\frac{\mu_1^2+r_{\perp j}^2}{z_j}\right]
                    \phi^{(n,n_1)}(\underline{q}_i,
                       \underline{r}_j,\underline{p})} \hspace{0.2in} \\
& =g&\left\{\sqrt{n+1}\int\frac{dq^+d^2q_\perp}{\sqrt{16\pi^3q^+}}
              \phi^{(n+1,n_1)}(\underline{q}_i,\underline{q},
                    \underline{r}_j,\underline{p}-\underline{q})\right.
\nonumber \\ 
& & +\frac{1}{\sqrt{n}}\sum_i\frac{1}{\sqrt{16\pi^3q_i^+}}
              \phi^{(n-1,n_1)}(\underline{q}_1,\ldots,\underline{q}_{i-1},
                      \underline{q}_{i+1},\ldots,\underline{q}_n,
                       \underline{r}_j,\underline{p}+\underline{q}_i)
\nonumber \\ 
& &+i\sqrt{n_1+1}\int\frac{dr^+d^2r_\perp}{\sqrt{16\pi^3r^+}}
              \phi^{(n,n_1+1)}(\underline{q}_i,\underline{r}_j,
                           \underline{r},\underline{p}-\underline{r})
\nonumber \\ 
& & +\left.\frac{i}{\sqrt{n_1}}\sum_j\frac{1}{\sqrt{16\pi^3r_j^+}}
              \phi^{(n,n_1-1)}(\underline{q}_i,\underline{r}_1,\ldots,
                                     \underline{r}_{j-1},
                        \underline{r}_{j+1},\ldots,\underline{r}_{n_1},
                           \underline{p}+\underline{r}_j) \right\}\,,
\nonumber 
\end{eqnarray} 
with $x=p^+/P^+$, $y_i=q_i^+/P^+$, and
$z_j=r_j^+/P^+$.


For fixed $M$, the eigenvalue problem itself is a condition on
the bare parameters.  A convenient choice for the second
condition is the value of an expectation value involving the
boson field $\phi(\underline{x})$; we use 
$\langle :\!\!\phi^2(0)\!\!:\rangle
  \equiv\Phi_\sigma^\dagger\!:\!\!\phi^2(0)\!\!:\!\Phi_\sigma$,
which corresponds to the expectation value for the sum of
$2/y_i$ for physical bosons.  For the soluble model in
Ref.~\cite{PV1} it was shown to be closely tied to the
coupling $g$, as can be seen in Eq.~(3.11) of that paper.
Most importantly, 
it can be computed rather quickly from a sum
similar to the normalization sum
\begin{eqnarray}
\langle :\!\!\phi^2(0)\!\!:\rangle
        =&\sum_{n=1,n_1=0}\prod_i^n&\int\,dq_i^+d^2q_{\perp i}
                     \prod_j^{n_1}\int\,dr_j^+d^2r_{\perp j}
                     \left(\sum_{k=1}^n \frac{2}{q_k^+/P^+}\right) \\
& &\times \left|\phi^{(n,n_1)}(\underline{q}_i,\underline{r}_j;
 \underline{P}-\sum_i\underline{q}_i-\sum_j\underline{r}_j)\right|^2\,.
    \nonumber
\end{eqnarray}
These two conditions are sufficient to fix $g$ and $M'_0$.

With the two parameters of the model now fully determined, we can compute
other quantities as predictions.  These are all obtained from the primary
output, which is the set of wave functions $\phi^{(n,n_1)}$ for the
different Fock sectors.  We will compute the slope of the no-flip form
factor of the fermion, structure functions for bosons and the fermion,
the distribution amplitude for the physical boson,
average momenta, average multiplicities, and a quantity sensitive to boson
correlations. The form factor slope $F'(0)$ is given by\cite{PV1} 
\begin{eqnarray} \label{eq:Fprime}
\lefteqn{F'(0)=\sum_{n,n_1}\prod_i^n\int\,dq_i^+d^2q_{\perp i}
           \prod_j^{n_1}\int\,dr_j^+d^2r_{\perp j}}\hspace{0.5in}\\
     & \times \left[\left(\sum_i \frac{y_i^2}{4}
                             \nabla_{\perp i}^2\right.\right. &
         +\left.\left. \sum_j \frac{z_j^2}{4}\nabla_{\perp j}^2\right)
                   \phi^{(n,n_1)}(\underline{q}_i,\underline{r}_j;
       \underline{P}-\sum_i\underline{q}_i-\sum_j\underline{r}_j)\right]^*
    \nonumber \\
    & &  \times
             \phi^{(n,n_1)}(\underline{q}_i,\underline{r}_j;
           \underline{P}-\sum_i\underline{q}_i-\sum_j\underline{r}_j)\,.
    \nonumber
\end{eqnarray}
A related form,
\begin{eqnarray} \label{eq:BetterFprime}
\lefteqn{\tilde{F}'(0)=-\sum_{n,n_1}\prod_i^n\int\,dq_i^+d^2q_{\perp i}
           \prod_j^{n_1}\int\,dr_j^+d^2r_{\perp j}}\hspace{0.5in} \\
     & \times & \left[\sum_i \left|\frac{y_i}{2}\nabla_{\perp i}
             \phi^{(n,n_1)}(\underline{q}_i,\underline{r}_j;
           \underline{P}-\sum_i\underline{q}_i-\sum_j\underline{r}_j)
                                     \right|^2 \right.
    \nonumber \\
    & & \left. +\sum_j \left|\frac{z_j}{2}\nabla_{\perp j}
                   \phi^{(n,n_1)}(\underline{q}_i,\underline{r}_j;
           \underline{P}-\sum_i\underline{q}_i
              -\sum_j\underline{r}_j)\right|^2 \right]\,,
    \nonumber
\end{eqnarray}
is better computationally.  It is obtained from (\ref{eq:Fprime})
via integration by parts.  If a momentum cutoff is present, there are
surface terms, but these will vanish at infinite cutoff.

The physical boson structure function is defined as
\begin{eqnarray}
f_B(y)\equiv&\sum_{n,n_1}\prod_i^n\int\,dq_i^+d^2q_{\perp i}
                     \prod_j^{n_1}\int&\,dr_j^+d^2r_{\perp j}
       \sum_{i=1}^n\delta(y-q_i^+/P^+) \\
   &&\times \left|\phi^{(n,n_1)}(\underline{q}_i,\underline{r}_j;
           \underline{P}-\sum_i\underline{q}_i
                             -\sum_j\underline{r}_j)\right|^2\,,
   \nonumber
\end{eqnarray}
The fermion and Pauli--Villars structure functions $f_F(x)$ and
$f_{PV}(z)$ are defined analogously.  The normalization of each
is such that the integral yields the average multiplicity
\begin{equation}
\langle n_B\rangle=\int_0^1f_B(y)dy\,,\;\;
\langle n_{PV}\rangle=\int_0^1f_{PV}(z)dz\,.
\end{equation}
The average momentum carried by each type is also given by
an integral
\begin{equation}
\langle y\rangle=\int_0^1yf_B(y)dy\,,\;\;
\langle z\rangle=\int_0^1zf_{PV}(z)dz\,.
\end{equation}
As a measure of the correlations in the multiple-boson Fock sectors,
we compute the covariance
$\langle y_1 y_2\rangle_{n\geq 2} -\langle y\rangle_{n\geq 2}^2$
where
\begin{eqnarray}
\langle y_1 y_2\rangle_{n\geq 2}=
  &\sum_{n\geq2,n_1}\prod_i^n\int\,dq_i^+d^2q_{\perp i}
                     \prod_j^{n_1}\int&\,dr_j^+d^2r_{\perp j}
       \sum_{i_1\neq i_2}^n \frac{q_{i_1}^+}{P^+}
                                     \frac{q_{i_2}^+}{P^+} \\
   &&\times \left|\phi^{(n,n_1)}(\underline{q}_i,\underline{r}_j;
           \underline{P}-\sum_i\underline{q}_i
                             -\sum_j\underline{r}_j)\right|^2\,,
   \nonumber
\end{eqnarray}
and $\langle y\rangle_{n\geq 2}$ is the same as $\langle y\rangle$
except that only states with two or more bosons are included.  We
also compute the distribution amplitude\cite{LepageBrodsky} given
by $\varphi(y)\equiv\int d^2q_\perp\phi^{(1,0)}(y,\bbox{q}_\perp)$.

\section{Numerical methods and results} \label{sec:NumMethods}

\subsection{Discretization and diagonalization}

We discretize the coupled integral equations and the
formulas for quantities such as the form factor slope
in the standard DLCQ manner\cite{PauliBrodsky}.
Integrals are approximated by discrete sums
and derivatives by finite-differences.  Because of the Pauli-Villars
regulation, the theory is ultraviolet finite.  However, in order to have a
finite matrix problem, we limit the range of transverse momentum by
imposing a cutoff $\Lambda^2$ on each constituent's invariant mass 
\begin{equation} 
\frac{m_i^2+p_{i\perp}^2}{x_i}\leq\Lambda^2\,,
\end{equation} 
where $m_i$ is the physical mass of the constituent. 
(Later, we study the large $\Lambda$ limit.) The longitudinal momentum,
always being positive, has a natural finite range.

Given the length scales $L$ and $L_\perp$, the discrete
momentum values are taken to be
\begin{equation}
p^+\rightarrow\frac{\pi}{L}n\,, \;\;
\bbox{p}_\perp\rightarrow
     (\frac{\pi}{L_\perp}n_x,\frac{\pi}{L_\perp}n_y)\,,
\end{equation}
with $n$ even for bosons and odd for fermions.
The differing values of $n$ correspond to use of periodic
and antiperiodic boundary conditions, respectively,
in a light-cone coordinate box
\begin{equation}
-L<x^-<L\,,\;\; -L_\perp<x,y<L_\perp\,.
\end{equation}
The total longitudinal momentum $P^+$ is used
to define an integer resolution\cite{PauliBrodsky}
$K\equiv\frac{L}{\pi}P^+$.
The positivity of the longitudinal integers $n$
implies that the number of particles in any Fock
sector is limited to $\sim\!\!K/2$.
The integers $n_x$ and $n_y$ range between limits associated with
some maximum integer $N_\perp$ fixed by $L_\perp$ and the
cutoff $\Lambda$, such that $N_\perp\pi/L_\perp$ is
the largest transverse momentum allowed by the cutoff.

The integral equations and other physical objects are
independent of $L$, a feature of boost-invariance in DLCQ.   The limit
$L\rightarrow\infty$ is replaced by the limit $K\rightarrow\infty$. The
momentum-space continuum limit is reached when both $K$ and $N_\perp$
become infinite. The momentum-space volume limit
$\Lambda^2\rightarrow\infty$ is taken after the continuum limit.

Weighting factors are included in the sums that
approximate integrals in order to incorporate boundary
effects induced by the invariant-mass cutoff.
For a discussion of how these factors are
constructed and used, see Ref.~\cite{PV1}.

Typical basis sizes are given in Table~\ref{tab:basis50}.
The present calculations, which use a single four-processor
node of an IBM SP, are limited to $\sim$11 million states.
The Hamiltonian matrix is extremely sparse, so that the
lowest-mass state can be efficiently extracted with use of
the Lanczos algorithm\cite{Lanczos} for complex symmetric matrices
\cite{Cullum,PV1}.  The analytic solution for the soluble
model discussed in Ref.~\cite{PV1} is used as an initial
guess for the Lanczos procedure.

\begin{table}
\mediumtext
\caption{\label{tab:basis50}
Basis sizes for DLCQ calculations
with parameters $M^2=\mu^2$, $\mu_1^2=10\mu^2$, and
$\Lambda^2=50\mu^2$.  The numbers of physical states
are in parentheses.}
\begin{tabular}{c|rrrrrrr}
  & \multicolumn{5}{c}{K} \\
 \cline{2-6}
$N_\perp$  &    9    &    11   &   13   &    15  &     17  \\
\hline
   5  &  54\,100 &   95\,176 &  386\,140 & 1\,553\,576 & 6\,816\,394  \\
      & (28\,065)&  (66\,371)& (232\,400)&(1\,038\,070) &(4\,972\,065) \\
   6  & 126\,748 &  536\,758 & 2\,907\,158 & 4\,935\,510 \\
      & (69\,245)& (391\,511)&(2\,107\,688) &(3\,013\,689) \\
   7  & 519\,325 & 1\,317\,392 & 10\,080\,748 \\
      &(276\,299)&(1\,008\,539) &(7\,272\,134) \\
   8  &1\,165\,832 & 5\,162\,002 \\
      &(687\,394) &(4\,140\,491) \\
   9  & 2\,268\,535 \\
      &(1\,437\,647)\\
  10  & 5\,850\,335\\
      &(3\,585\,752)\\
\end{tabular}
\narrowtext
\end{table}

Before invoking the Lanczos algorithm, the eigenvalue problem
is rearranged so that $-1/g$ is the eigenvalue.  This
allows computation of $g$ given a fixed value for $M$ and a guess for
$M'_0$.  The iterative Brent--M\"uller algorithm\cite{DeVries} 
is then used to find the value of $M'_0$ that brings 
$\langle:\!\!\phi^2(0)\!\!:\rangle$ into agreement with 
its chosen value.

\subsection{Results}

Most of the calculations reported here use the parameter values
$M^2=\mu^2$ and $\mbox{$\langle :\!\!\phi^2(0)\!\!:\rangle$}=1$.  
These choices correspond
to a relativistic, weak-coupling regime.   Because of the weak coupling,
the number of Fock sectors can be truncated to include no more than four
bosons without any discernible effect, as can be seen from the Fock-sector
probabilities listed in Table~\ref{tab:FockSectorProb}; most weak-coupling
calculations were done with this truncation in order to increase the
available momentum resolution. For comparison, we have also done some
study of other regimes.

\begin{table}
\caption{\label{tab:FockSectorProb}
Fock sector probabilities
$\int|\phi^{(n,n_1)}|^2\prod_i^nd\underline{q}_i
                               \prod_j^{n_1}d\underline{r}_j$, 
where $n$ is the number of physical bosons and $n_1$ the number of
Pauli--Villars bosons. The numerical and physical parameters are $K=17$,
$N_\perp=7$, $M^2=\mu^2$, $\mu_1^2=10\mu^2$, $\Lambda^2=50\mu^2$, and
$\langle:\!\!\phi^2(0)\!\!:\rangle=1$. The total number of bosons $n+n_1$
is limited to a maximum of 4. Probabilities smaller than $\sim10^{-5}$ are
not resolved with any accuracy.} 
\begin{tabular}{l|ccccc} 
$n\backslash n_1$ & 0 & 1 & 2 & 3 & 4 \\ 
\hline
0 & 0.8515 & 0.0115 & 0.8$\cdot 10^{-5}$ &
                  $\sim 10^{-10}$ & $\sim 10^{-16}$ \\
1 & 0.1333 & 0.0005 & $\sim 10^{-7}$ & $\sim 10^{-12}$ \\
2 & 0.0036 & 0.4$\cdot 10^{-5}$ & $\sim 10^{-10}$ \\ 
3 & 0.3$\cdot 10^{-4}$ & $\sim 10^{-8}$ \\ 
4 & $\sim 10^{-7}$  \\
\end{tabular}
\end{table}

Table~\ref{tab:extrapolated}
shows values of various quantities,
extrapolated from longitudinal
resolutions $K=9$ to 19 (or even 21) and transverse
resolutions $N_\perp=5$ to 10 for small $K$ and to 6 or 7
for large $K$.  These include
the bare coupling $g$, the renormalization parameter $M'_0$,
the bare fermion probability $|\psi_0|^2$,
the slope of the form factor $F'(0)$,
the average multiplicity $\langle n_B \rangle$, and
a parameterization of the structure function
$f_B(y)=Ay^a(1-y)^b$ (which is an excellent fit).
Each is shown as a function of the cutoff $\Lambda^2$
and the Pauli--Villars mass $\mu_1$.
The extrapolations were done by fitting to the form
$\alpha +\beta/K^2 +\gamma/N_\perp^2$; most
quantities are slowly varying with respect to
resolution.  The range of values obtained for $F'(0)$
correspond to a dressed-fermion radius $\sqrt{-6F'(0)}$
on the order of $0.2\,\mu^{-1}$.

\begin{table}
\caption{\label{tab:extrapolated}
Extrapolated bare parameters and observables.
The physical parameter
values were $M^2=\mu^2$ for the fermion mass
and $\langle:\!\!\phi^2(0)\!\!:\rangle=1$.}
\begin{tabular}{r|ccc|cc|cc}
 & \multicolumn{3}{c|}{$\mu_1^2=5\mu^2$}
 & \multicolumn{2}{c|}{$\mu_1^2=10\mu^2$}
 & \multicolumn{2}{c}{$\mu_1^2=20\mu^2$} \\
\hline
$(\Lambda/\mu)^2$ & 12.5 & 25 & 50 & 25 & 50 & 50 & 100 \\
\hline
$g/\mu$           & 21.4  & 17.7  & 16.3  & 17.8  & 16.0  & 16.0  & 15.5 \\
$M'_0/\mu^2$      & 1.26  & 1.10  & 1.10  & 1.48  &  1.4  & 1.8   & 1.9 \\
\hline
$|\psi_0|^2$      & 0.82  & 0.83  & 0.84  & 0.85  & 0.86  & 0.87  & 0.87 \\
$-100\mu^2\tilde{F}'(0)$ 
                  & 1.04  & 0.78  & 0.66  & 0.72  & 0.59  & 0.59  & 0.51 \\
$\langle n_B\rangle$ 
                  & 0.18  & 0.15  & 0.14  & 0.15  & 0.14  & 0.13  & 0.13 \\
$\langle y\rangle$& 0.077 & 0.062 & 0.057 & 0.062 & 0.056 & 0.056 & 0.053 \\
$\langle y_1y_2\rangle_{n\geq 2}-\langle y\rangle_{n\geq 2}^2$
                  & $1.1\cdot10^{-3}$ & $6\cdot10^{-4}$ & $6\cdot10^{-4}$
                      & $6\cdot10^{-4}$ & $6\cdot10^{-4}$ & $6\cdot10^{-4}$
                      & $5\cdot10^{-4}$ \\
$A$               & 9.39 & 4.21 & 3.00 & 4.15 & 2.77 & 2.7 & 2.4 \\ 
$a$               & 1.90 & 1.50 & 1.36 & 1.48 & 1.31 & 1.29 & 1.26 \\ 
$b$               & 2.95 & 2.54 & 2.32 & 2.53 & 2.26 & 2.24 & 2.14 \\ 
\end{tabular} 
\end{table}

The table shows that the renormalization parameter $M'_0$ is the only
quantity  strongly dependent on the Pauli--Villars mass.  This is to 
be expected because of its role in the self-energy counterterm.
One might argue that $F'(0)$ is also strongly dependent; however, 
any apparent variation with $\mu_1^2$ is largely due to differences 
in cutoff values and transverse resolution.  Although $F'(0)$ will 
ultimately become independent of $\Lambda^2$ and $N_\perp$, it is 
sensitive to these in the ranges where we calculate.
The table also shows that the estimate of $\sum_i y_i<<1$ by G{\l}azek
and Perry\cite{GlazekPerry} is justified, in that the expectation 
value $\langle y\rangle$ is found to be small.

\begin{figure}[p]
\centerline{\epsfxsize=\columnwidth \epsfbox{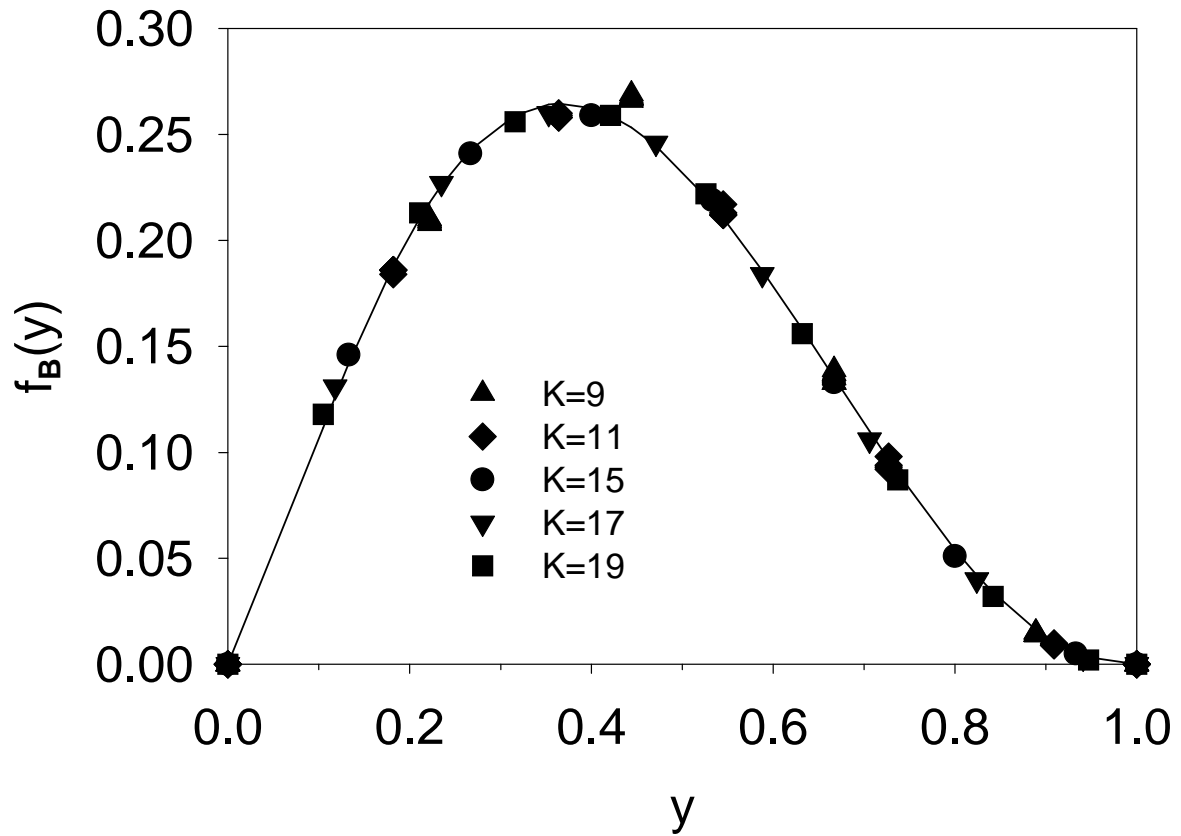} }
\caption{\label{fig:fB} The boson structure function $f_B$ at various
numerical resolutions, with $M=\mu$, $\langle:\!\!\phi^2(0)\!\!:\rangle=1$,
$\Lambda^2=50\mu^2$, and $\mu_1^2=10\mu^2$. The solid line is the
parameterized fit, $Ay^a(1-y)^b$.} 
\end{figure}

\begin{figure}[p]
\centerline{\epsfxsize=\columnwidth \epsfbox{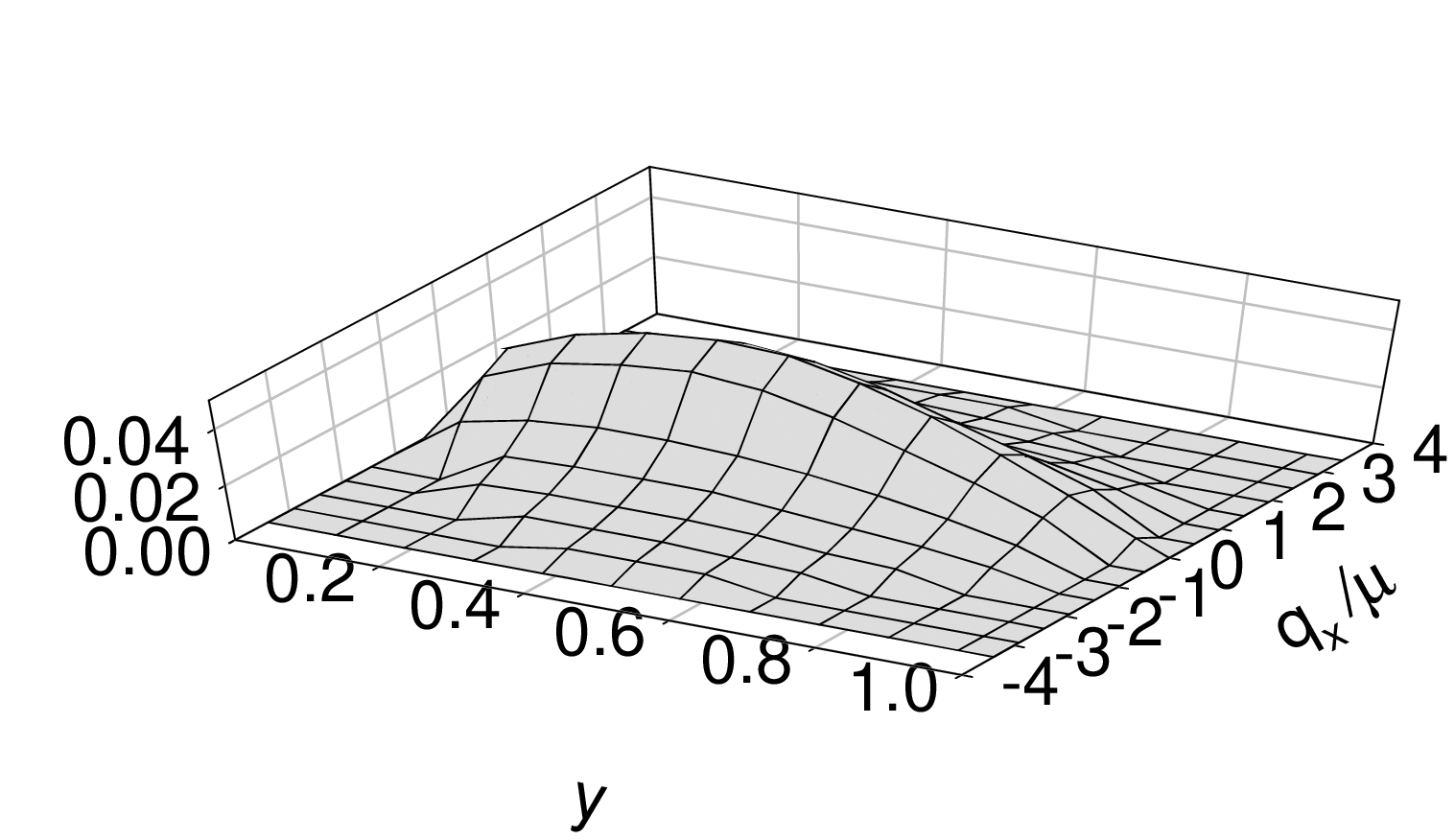} }
\caption{\label{fig:DynamicalAmplitude}
The one-boson amplitude $\phi^{(1,0)}$ as a function of longitudinal
momentum fraction $y$ and one transverse momentum component $q_x$ in the
$q_y=0$ plane. The parameter values are $K=21$, $N_\perp=7$,
$\mu_1^2=10\mu^2$, $\Lambda^2=25\mu^2$, and
$\langle:\!\!\phi^2(0)\!\!:\rangle=1$. } 
\end{figure}

\begin{figure}[p]
\centerline{\epsfxsize=\columnwidth \epsfbox{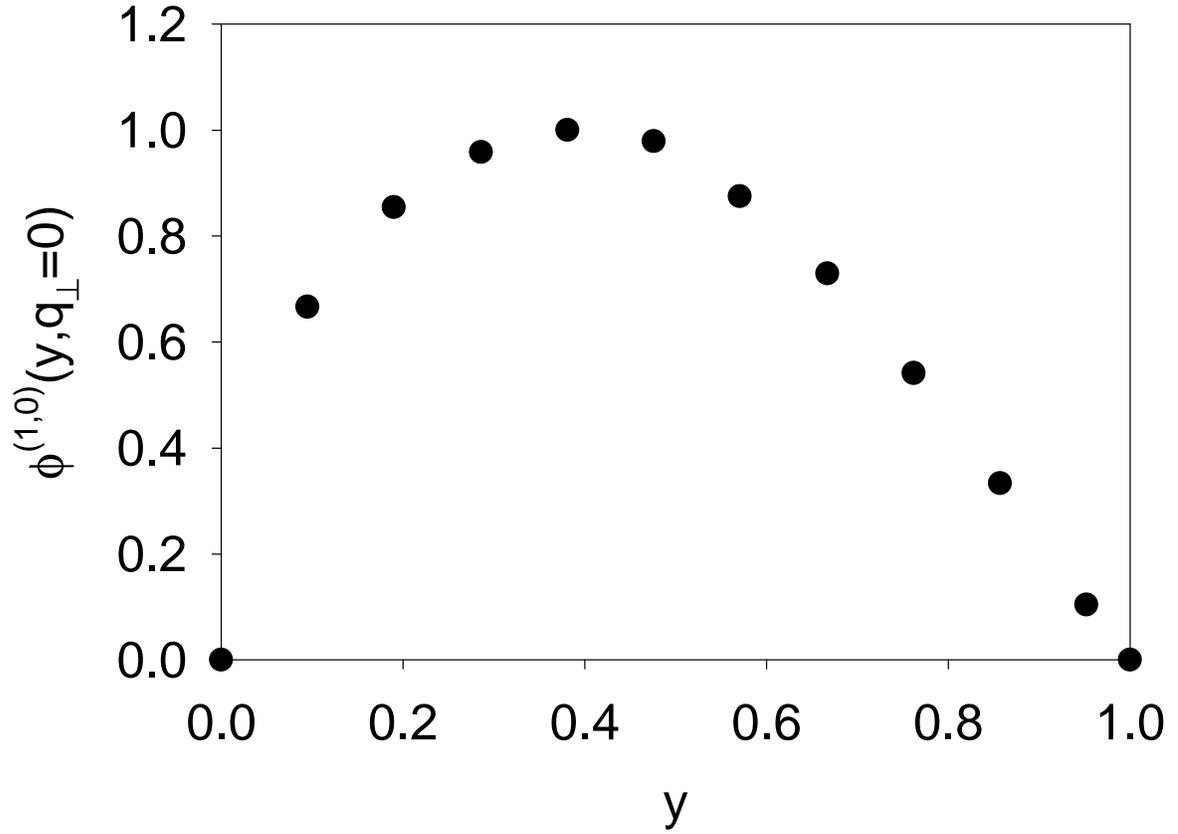} }
\caption{\label{fig:twoblong}
The boson-fermion two-body amplitude at zero transverse
momentum, with $K=21$,
$N_\perp=7$,
$\langle:\!\!\phi^2(0)\!\!:\rangle=1$,
$\Lambda^2=25\mu^2$, and $\mu_1^2=10\mu^2$.
The normalization is arbitrary.}
\end{figure}

\begin{figure}[p]
\centerline{\epsfxsize=\columnwidth \epsfbox{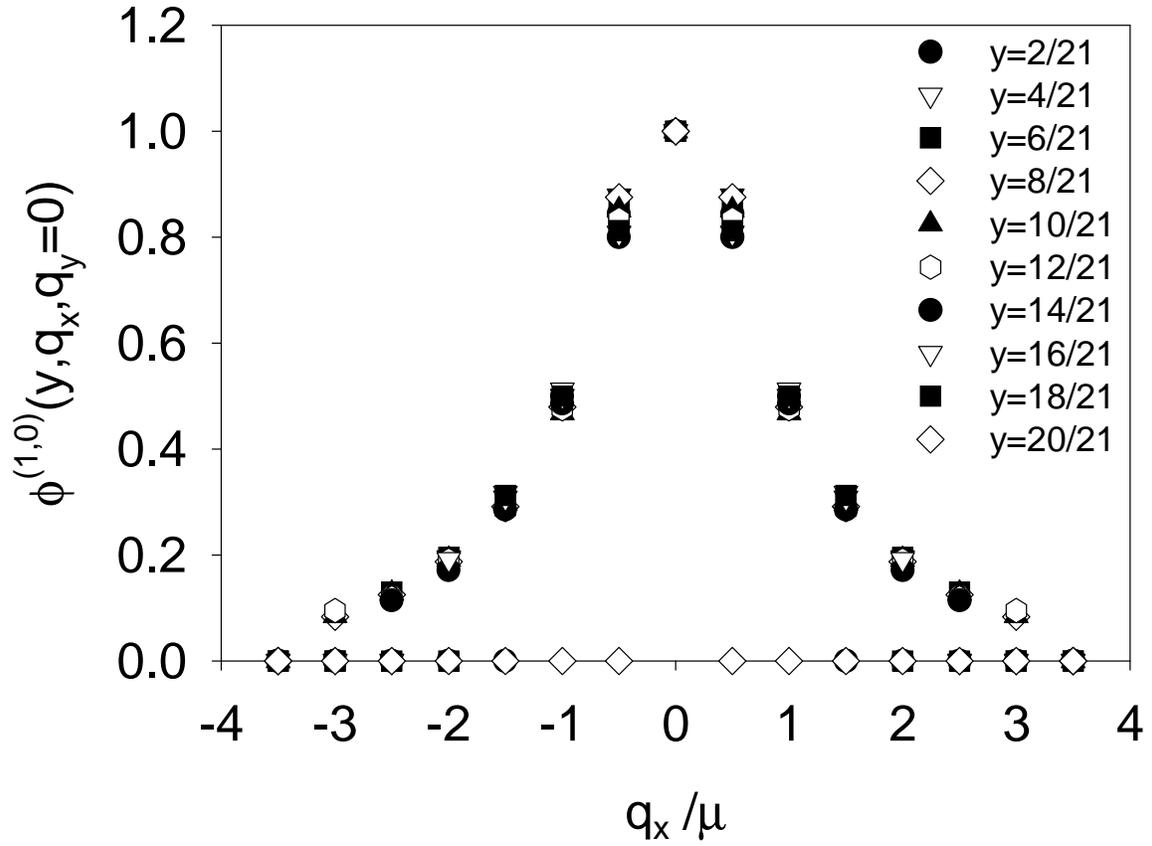} }
\caption{\label{fig:twobody}
Cross sections of the boson-fermion two-body amplitude
taken at varying longitudinal momenta and at fixed $q_y=0$,
with $K=21$, $N_\perp=7$, $\langle:\!\!\phi^2(0)\!\!:\rangle=1$,
$\Lambda^2=25\mu^2$, and $\mu_1^2=10\mu^2$.
The peaks are normalized to be equal at $q_x=0$.}
\end{figure}

\begin{figure}[p]
\centerline{\epsfxsize=\columnwidth \epsfbox{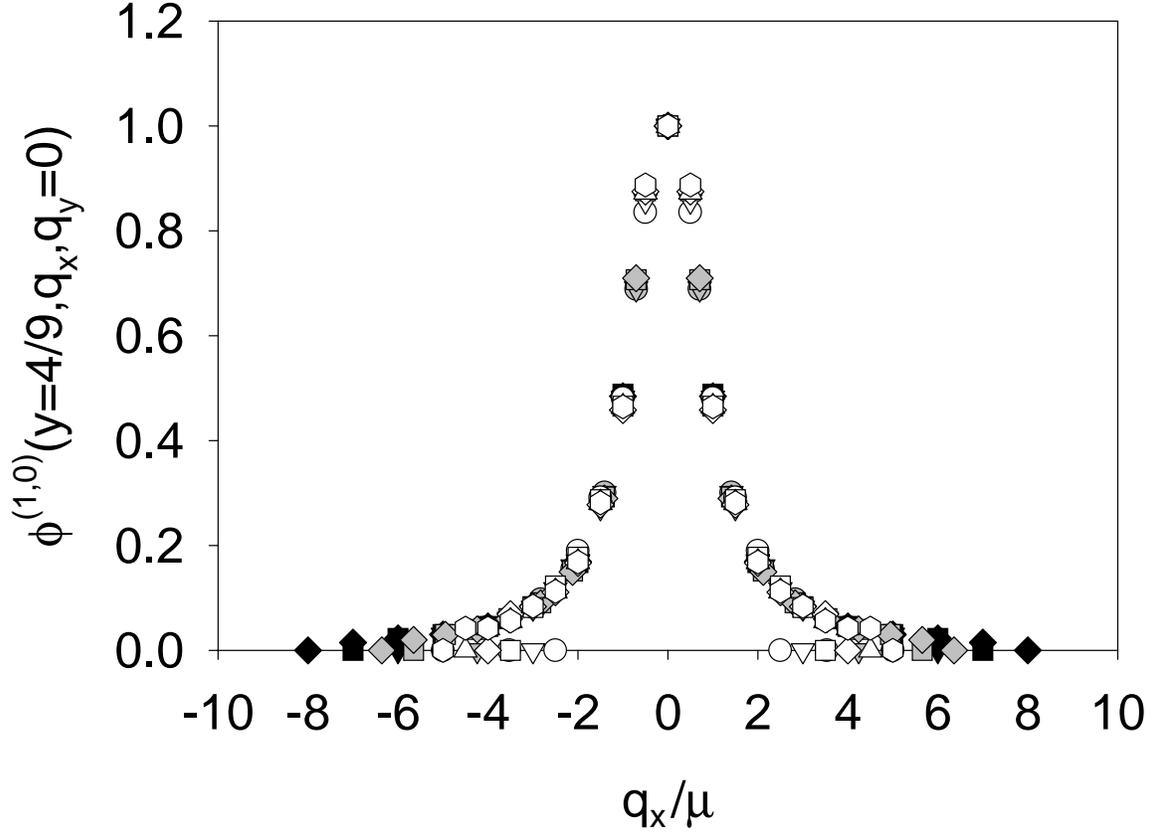} }
\caption{\label{fig:fixedlp}
A cross section of the boson-fermion two-body amplitude
taken at fixed longitudinal momentum fraction $y=4/9$
and at fixed $q_y=0$, with $K=9$,
$\langle:\!\!\phi^2(0)\!\!:\rangle=1$,
and $\mu_1^2=10\mu^2$.  The cutoff $\Lambda^2$ and
the transverse resolution $N_\perp$ are varied to
keep the transverse scale $L_\perp$ fixed
at one of the following values: $1\frac{\pi}{\mu}$ (black),
$\sqrt{2}\frac{\pi}{\mu}$ (gray),
and $2\frac{\pi}{\mu}$ (white).  Different symbol shapes
correspond to different values of $N_\perp$.
The peaks are normalized to be equal at $q_x=0$.
The points at zero amplitude mark the transverse range, which is
set by the cutoff.}
\end{figure}

\begin{figure}[p]
\centerline{\epsfxsize=\columnwidth \epsfbox{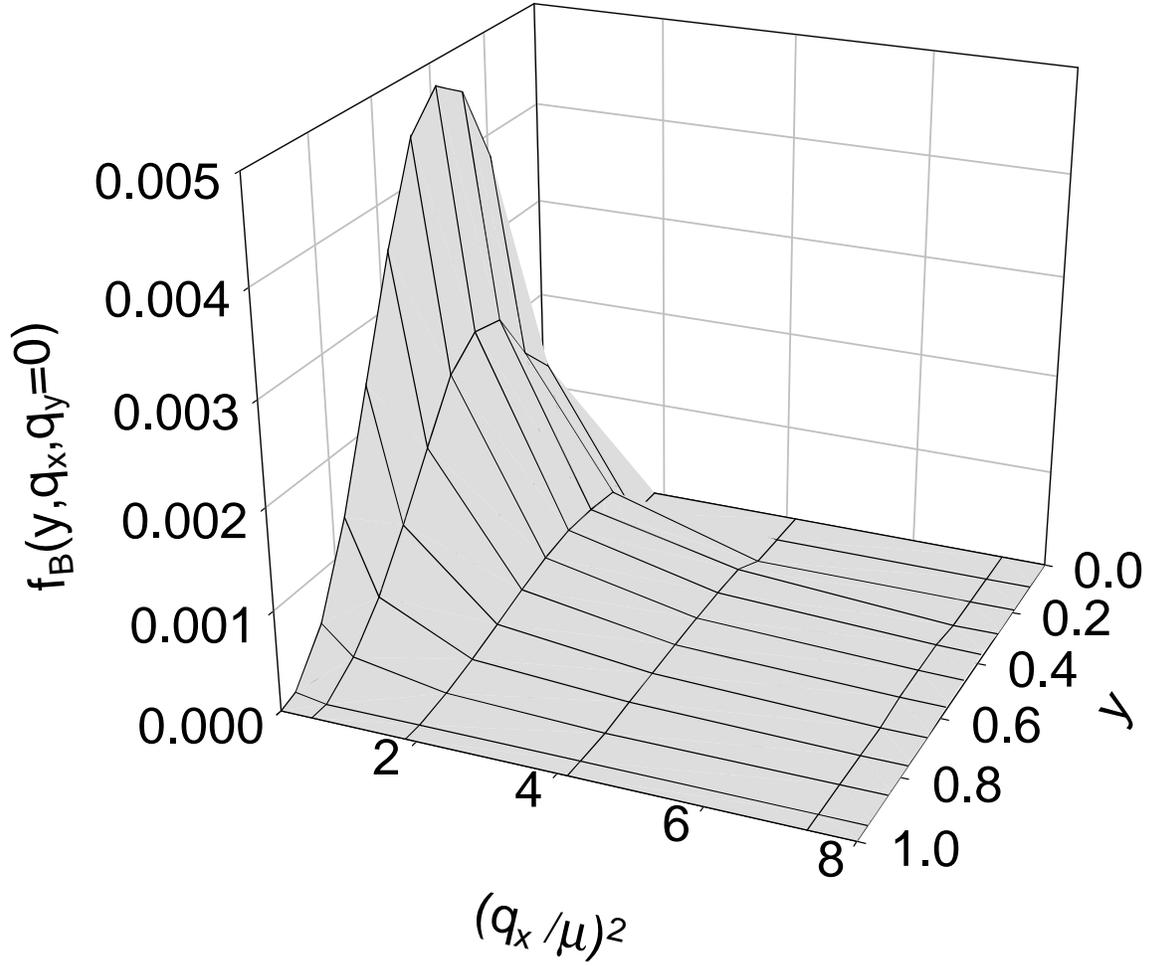} }
\caption{\label{fig:fbq2} 
The boson structure function $f_B(y,{\bf q}_\perp)$ 
with $K=21$, $N_\perp=7$,
$\langle:\!\!\phi^2(0)\!\!:\rangle=1$, $\Lambda^2=25\mu^2$, and
$\mu_1^2=10\mu^2$. The transverse momentum is varied with 
$q_y$ fixed at zero. } 
\end{figure}

\begin{figure}[p]
\centerline{\epsfxsize=\columnwidth \epsfbox{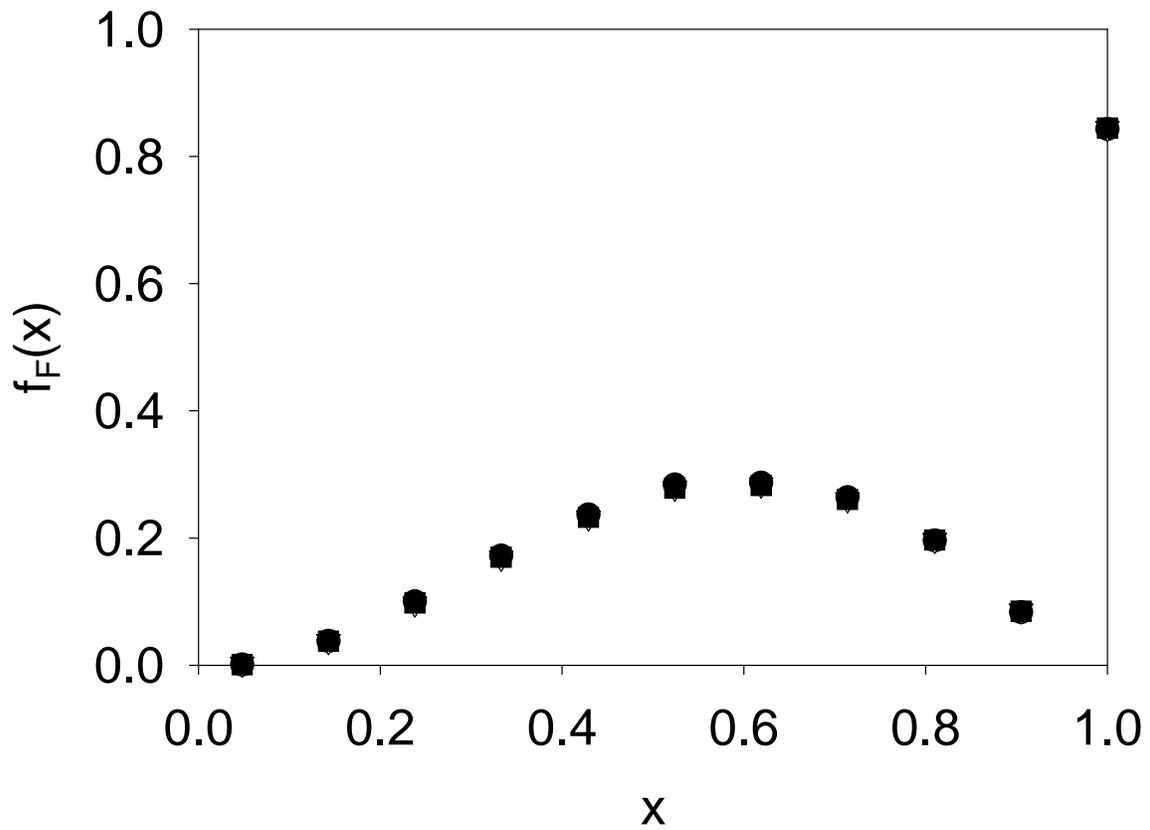} }
\caption{\label{fig:fF} The fermion structure function $f_F$ with
$K=21$, $N_\perp=5$ to 7, $\langle:\!\!\phi^2(0)\!\!:\rangle=1$,
$\Lambda^2=25\mu^2$, and $\mu_1^2=10\mu^2$. Each $N_\perp$ value yields
essentially the same result.} 
\end{figure}

\begin{figure}[p]
\centerline{\epsfxsize=\columnwidth \epsfbox{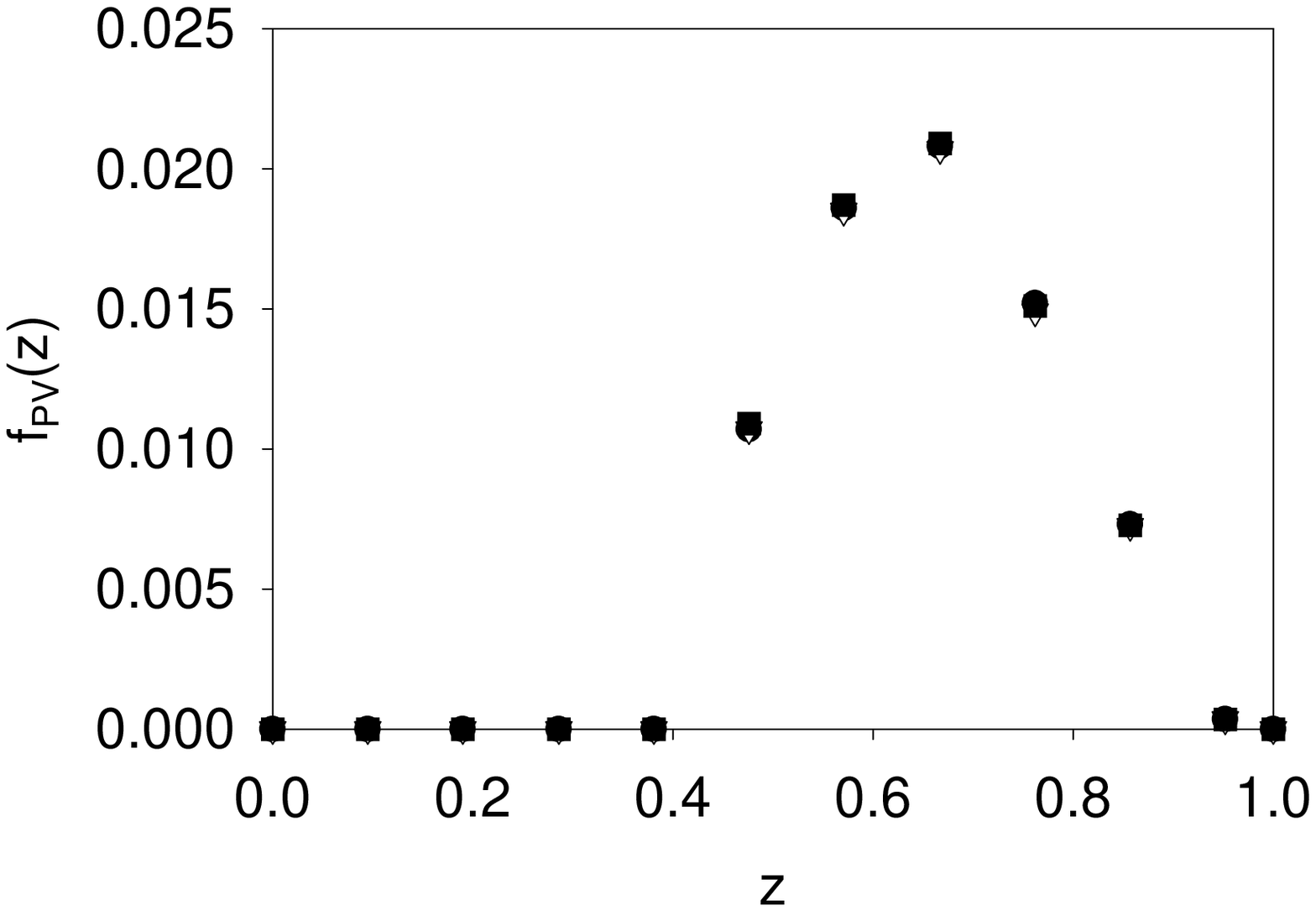} }
\caption{\label{fig:fPV} Same as Fig.~\ref{fig:fF} but for the
Pauli--Villars boson structure function $f_{PV}$.} 
\end{figure}

A sample boson structure function is plotted in Fig.~\ref{fig:fB}.  The
figure also shows how well the form $Ay^a(1-y)^b$ fits the numerical
results and how insensitive $f_B$ is to numerical resolution,
something which was also observed for the model considered in
Ref.~\cite{PV1}. The transverse and longitudinal dependence of a two-body
amplitude are shown in Figs.~\ref{fig:DynamicalAmplitude},
\ref{fig:twoblong} and \ref{fig:twobody}.  A particular transverse cross
section of the two-body amplitude is presented in Fig.~\ref{fig:fixedlp};
these results correspond to fixed values of the transverse scale $L_\perp$
and are remarkably consistent. Figure~\ref{fig:fbq2} shows the $Q^2$
dependence of the boson structure function. 
A fermion structure function and
a Pauli--Villars boson structure function are plotted in
Figs.~\ref{fig:fF} and \ref{fig:fPV}.  The parameter values are the same
for both.  The skewing of the Pauli-Villars particle momentum
distributions to high longitudinal momentum fractions reflects the heavy
mass of the Pauli--Villars bosons.

\begin{figure}[p]
\centerline{\epsfxsize=\columnwidth \epsfbox{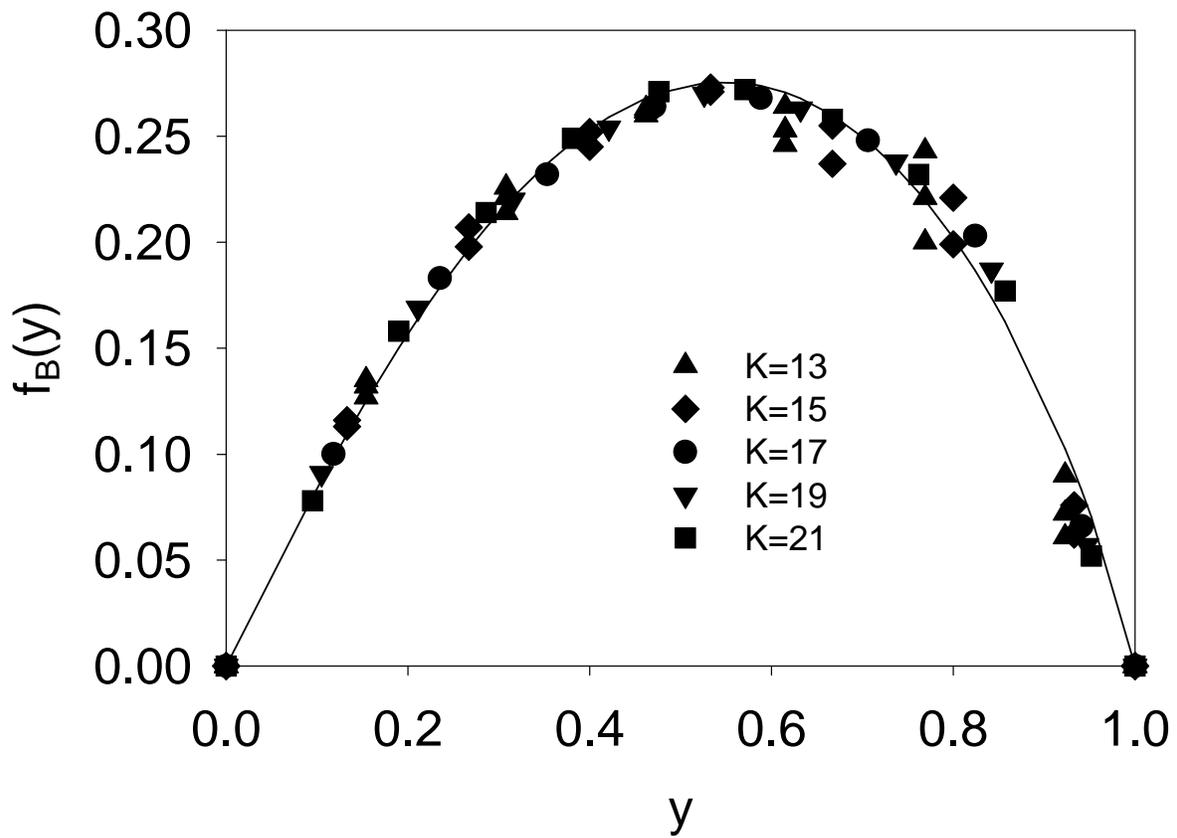} }
\caption{\label{fig:fBm01} Same as Fig.~\protect\ref{fig:fB},
but for $M^2=0.1\mu^2$.} 
\end{figure}

\begin{figure}[p]
\centerline{\epsfxsize=\columnwidth \epsfbox{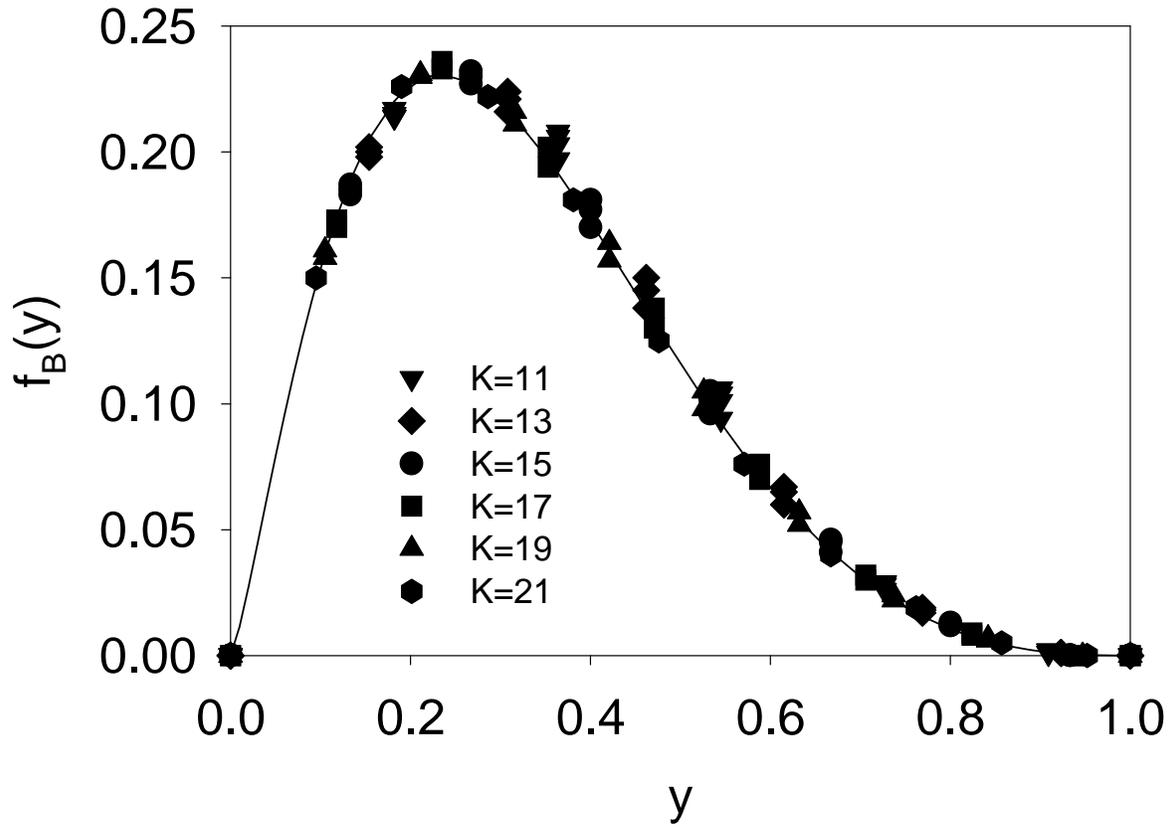} }
\caption{\label{fig:fBm5} Same as Fig.~\protect\ref{fig:fB},
but for $M^2=5\mu^2$ and $\Lambda=100\mu^2$
and a parameterized fit of $Ay^a(1-y)^be^{-cy}$.} 
\end{figure}

\begin{figure}[p]
\centerline{\epsfxsize=\columnwidth \epsfbox{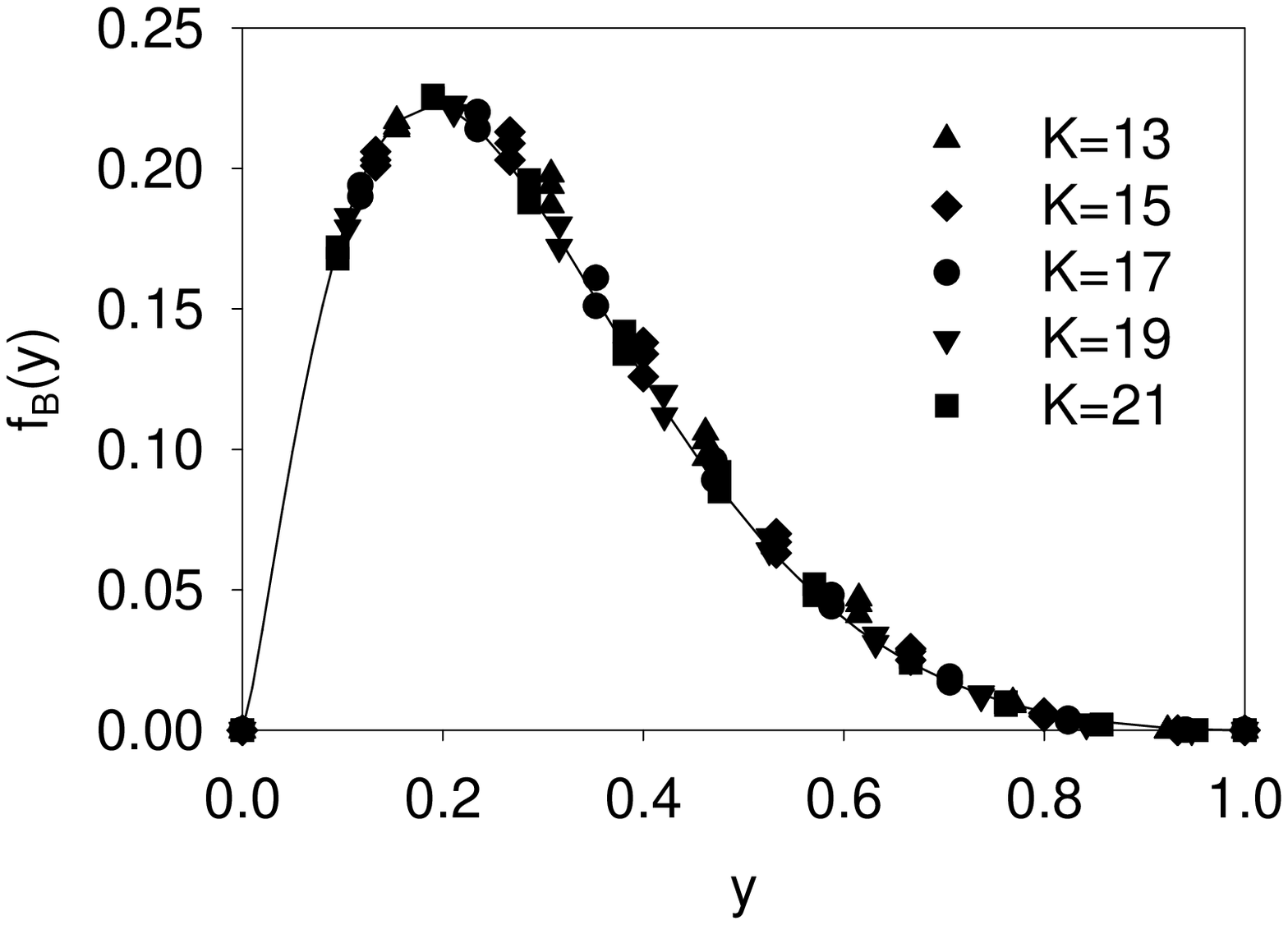} }
\caption{\label{fig:fBm10} Same as Fig.~\protect\ref{fig:fBm5},
but for $M^2=10\mu^2$.} 
\end{figure}

\begin{figure}[p]
\centerline{\epsfxsize=\columnwidth \epsfbox{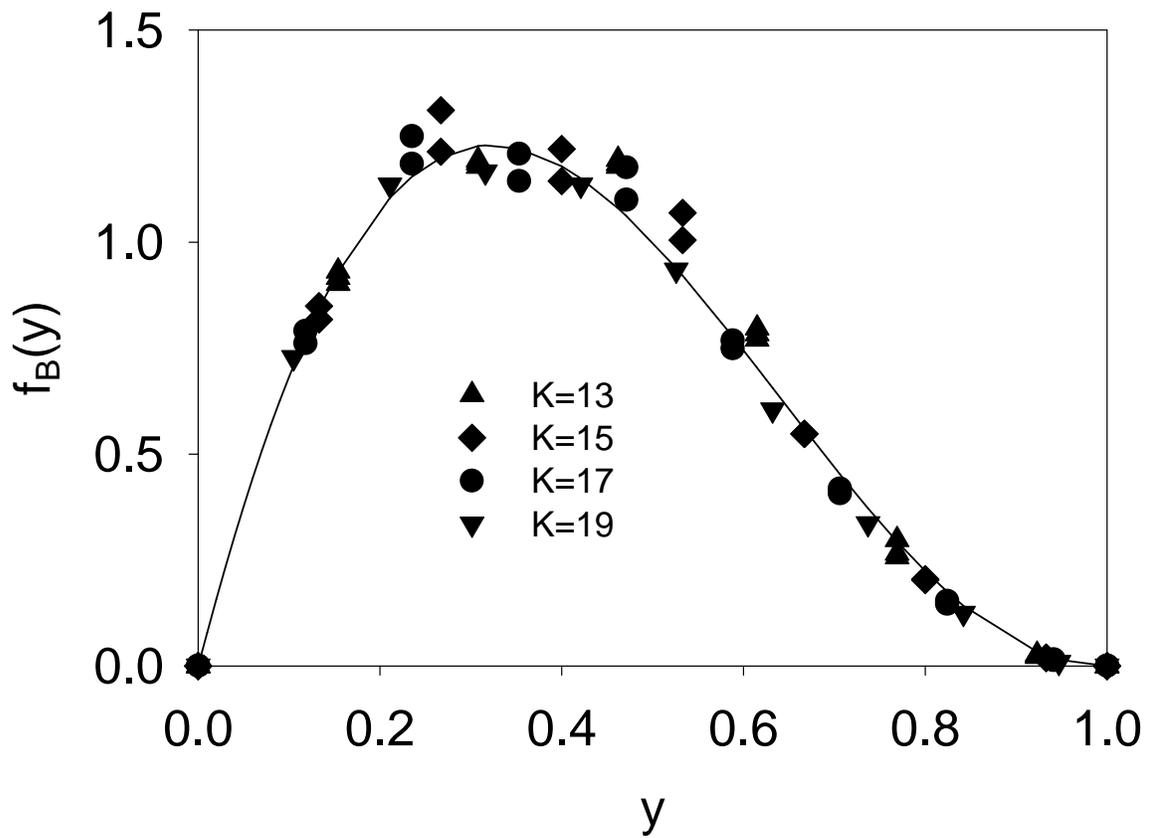} }
\caption{\label{fig:fBp5} Same as Fig.~\protect\ref{fig:fB},
but for $\langle:\!\!\phi^2(0)\!\!:\rangle=5$.} 
\end{figure}

\begin{figure}[p]
\centerline{\epsfxsize=\columnwidth \epsfbox{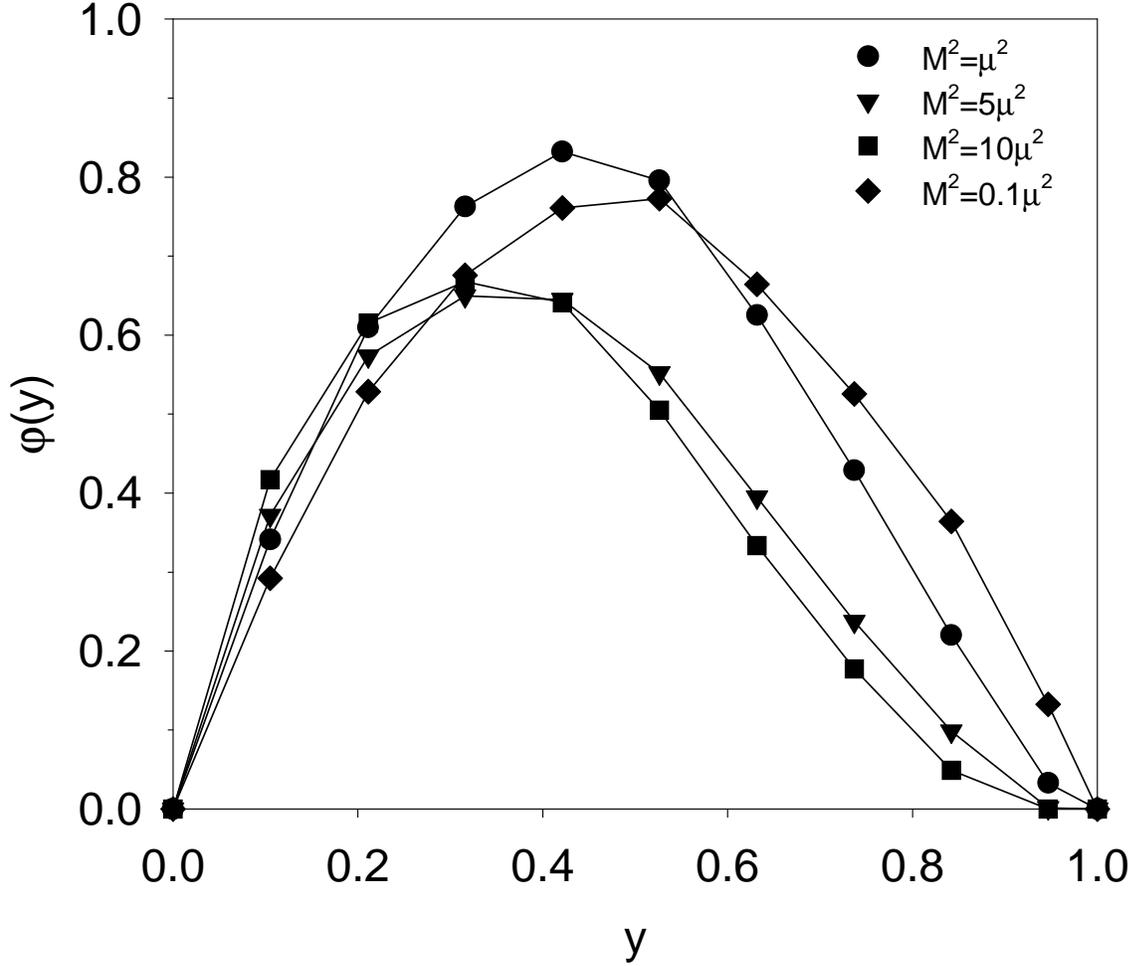} }
\caption{\label{fig:Phi-M} Comparison of distribution
amplitudes $\varphi(y)\equiv\int d^2 q_\perp \phi^{(1,0)}(y,\bbox{q}_\perp)$.
Various values are considered for the fermion mass $M$, with
$\langle:\!\!\phi^2(0)\!\!:\rangle=1$.  The values of the numerical
parameters are $K=19$, $N_\perp=5$, $\Lambda^2=50\mu^2$ (except
for $M^2=5\mu^2$ and $10\mu^2$ when $\Lambda^2=100\mu^2$), and
$\mu_1^2=10\mu^2$.  The lines simply connect the computed points,
to guide the eye, and the absolute normalization is arbitrary.} 
\end{figure}

\begin{figure}[p]
\centerline{\epsfxsize=\columnwidth \epsfbox{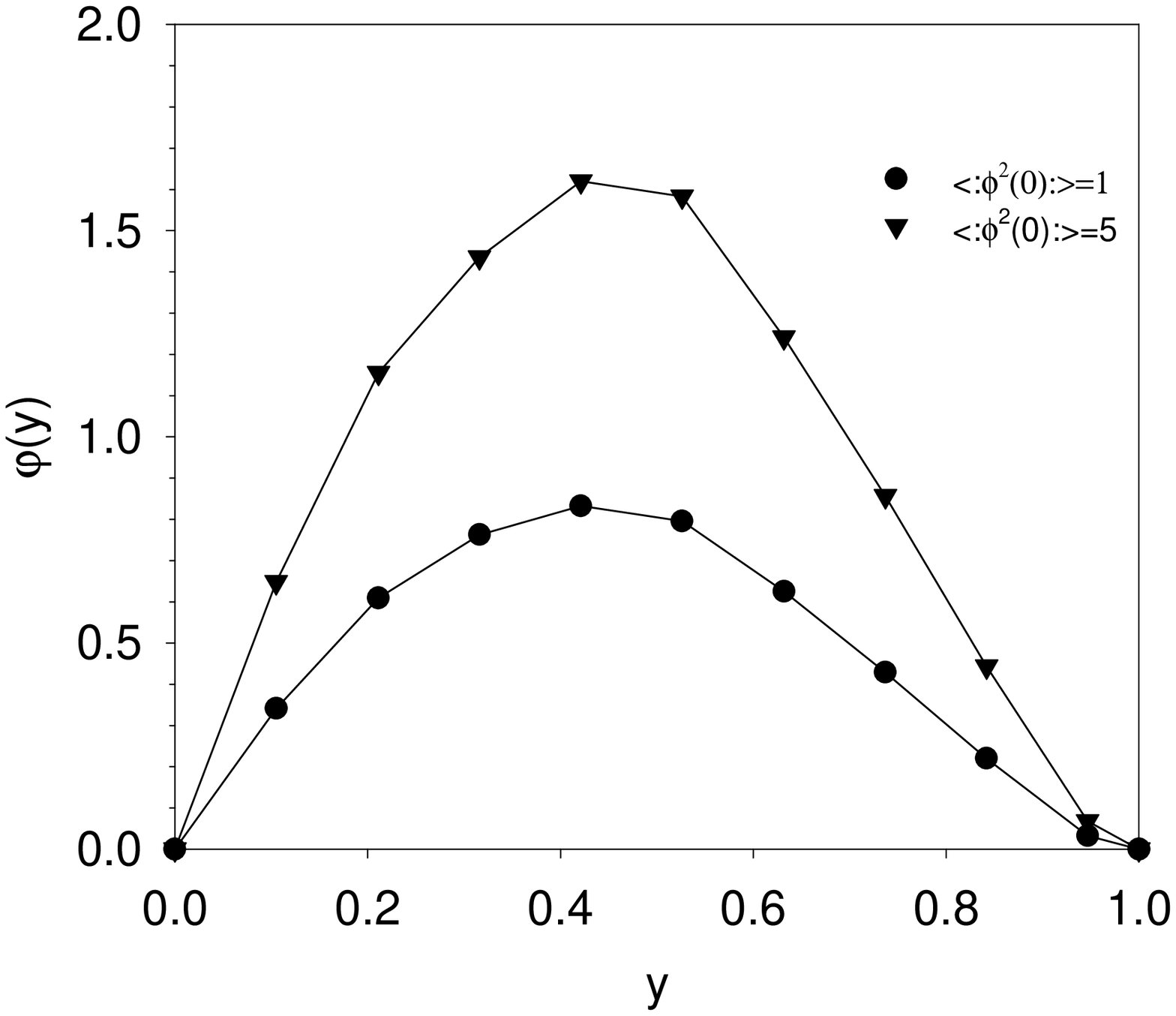} }
\caption{\label{fig:Phi-P} 
Same as Fig.~\protect\ref{fig:Phi-M} but with the fermion
mass fixed at $M=\mu$ and $\langle:\!\!\phi^2(0)\!\!:\rangle$
varied.} 
\end{figure}

Other values for the physical parameters $M$ and 
$\langle:\!\!\phi^2(0)\!\!:\rangle$ have also been
considered.  A summary of extrapolated quantities
is given in Table~\ref{tab:extrapII}.  The associated
structure functions $f_B(y)$ are shown in Figs.~\ref{fig:fBm01}
through \ref{fig:fBp5}.  Distribution amplitudes are
displayed in Figs.~\ref{fig:Phi-M} and \ref{fig:Phi-P}.
For values of $M$ larger than
$\mu$ we have found the form $Ay^a(1-y)^be^{-cy}$ to
allow a noticeably better fit to $f_B(y)$.
For $\langle:\!\!\phi^2(0)\!\!:\rangle=5$
the maximum number of bosons was increased to 5.
The numerical resolutions ranged from 9 to 21 for
$K$ and from 5 to as much as 10 for $N_\perp$.

The extent to which the fermion source is dressed by the bosons
is directly determined by the mass ratio $M/\mu$ and the
coupling strength.  The latter is tightly correlated with
the chosen observable $\langle:\!\!\phi^2(0)\!\!:\rangle$.
As the ratio $M/\mu$ is tuned, the boson structure function
$f_B(y)$ shifts dramatically.  A relatively small boson
mass shifts the peak in $f_B(y_)$ to small boson momentum
fractions, as shown in Fig.~\ref{fig:fBm10}.  A large 
mass shifts the peak to central values of $y$ and
significantly raises the constituent density at
large $y$, as illustrated in Fig.~\ref{fig:fBm01}.
An increase in $\langle:\!\!\phi^2(0)\!\!:\rangle$
increases the coupling and increases the probability
for a large number of constituents.  Analogous changes
occur for the distribution amplitude.  Comparison
of Tables~\ref{tab:extrapolated} and \ref{tab:extrapII}
shows that the average number $\langle n_B\rangle$
increases significantly when $\langle:\!\!\phi^2(0)\!\!:\rangle$
is changed from 1 to 5.

\begin{table}
\caption{\label{tab:extrapII}
Same as Table~\protect\ref{tab:extrapolated}, but
for different $M^2$ or $\langle:\!\!\phi^2(0)\!\!:\rangle$
values.}
\begin{tabular}{r|ccc|c}
 & \multicolumn{3}{c|}{$\langle:\!\!\phi^2(0)\!\!:\rangle=1$}
 & $\langle:\!\!\phi^2(0)\!\!:\rangle=5$ \\
\hline
$(M/\mu)^2$       & 0.1 & 5 & 10 & 1 \\
$(\mu_1/\mu)^2$   & 10 & 10 & 10 & 10  \\
$(\Lambda/\mu)^2$ & 50 & 100 & 100 & 50  \\
\hline
$g/\mu$           & 15.1 & 18.1 & 19.0 & 44.5 \\
$M'_0/\mu^2$      & 1.39 & 1.66 & 1.60 & 10.1 \\
\hline
$|\psi_0|^2$      & 0.83 & 0.89 & 0.90 & 0.41 \\
$-100\mu^2\tilde{F}'(0)$ 
                  & 2.0 & 0.14 & 0.07 & 6.7  \\
$\langle n_B\rangle$ 
                  & 0.16 & 0.10 & 0.09 & 0.62 \\
$\langle y\rangle$& 0.073 & 0.032 & 0.024 & 0.24 \\
$\langle y_1y_2\rangle_{n\geq 2}-\langle y\rangle_{n\geq 2}^2$
                  & $7\cdot 10^{-4}$ & $3\cdot 10^{-4}$ & 
                     $3\cdot 10^{-4}$ & $8\cdot 10^{-3}$ \\
$A$               &  1.0333 & 5.2548 & 7.5519 & 9.0847 \\ 
$a$               &  1.0512 & 1.3191 & 1.3339 & 1.0256 \\ 
$b$               &  0.8678 & 2.5430 & 1.7151 & 2.1580 \\
$c$               &  0      & 2.2730 & 4.9870 & 0 \\ 
\end{tabular} 
\end{table}

\section{Conclusion}  \label{sec:Conclusions}

We have successfully computed the Fock-sector wave functions
which fully describe the lowest-mass eigenstate of a field-theoretic model
Hamiltonian (\ref{eq:ModelH}) in physical three space and one time
dimensions.  From these wave functions we have
extracted several interesting quantities to show that numerical
convergence is under control and that Pauli--Villars regularization leads
to sensible results.  The size of the momentum-state basis required is
large but manageable for present-day computing machines.  Larger bases
could be used by expanding to more than one node, although one then pays
the price of message-passing overhead.

For the model discussed here there are still interesting
calculations which might be done.  One could look at excited
states in the one-fermion sector that we have explored,
or consider other sectors, such as the two-fermion sector.
Extension to two flavors, particularly with very different
masses, should yield some understanding of light
systems with heavy intrinsic constituents, which could
have some relevance for intrinsic charm\cite{IntrinsicCharm}.

Beyond this model there are, of course, many possibilities.
A solution of Yukawa theory\cite{YukawaLFTD}, in a no-pair
approximation or eventually in full, would be the most
immediate nontrivial extension.  Applications to quantum
electrodynamics, to positronium\cite{TangPauli} or the
electron's anomalous moment\cite{ae} in particular,
would be quite natural.  Direct application to quantum
chromodynamics (QCD) may be problematic; however, a
supersymmetric conformally-invariant form of QCD 
could lend itself to the spirit of the approach, in 
that heavy superpartners in a broken supersymmetry 
should provide the needed ultraviolet cancellations.

\acknowledgments
This work was supported in part by the Minnesota Supercomputer Institute
through grants of computing time and by the Department of Energy,
contracts DE-AC03-76SF00515 (S.J.B.),  DE-FG02-98ER41087 (J.R.H.), and
DE-FG03-95ER40908 (G.M.). The hospitality of the Aspen Center for Physics
was also appreciated.

\end{document}